%% file: SDCL.tex
\documentclass[runningheads]{llncs}
\usepackage{graphicx}
\usepackage{hyperref}
\hypersetup{colorlinks=true,allcolors=blue}
\usepackage{adjustbox}
\usepackage{cite}
\usepackage{marvosym}

\begin{document}

\title{
SDCL: Students Discrepancy-Informed Correction Learning for Semi-supervised Medical Image Segmentation
}

\titlerunning{SDCL: Students Discrepancy-Informed Correction Learning}

\author{Bentao Song\index{Song,Bentao} \and
Qingfeng Wang\index{Wang,Qingfeng} $^{(\textrm{\Letter})}$}

\authorrunning{B. Song et al.}

\institute{School of Computer Science and Technology, Southwest University of Science and Technology, China
\\\email{qfwangyy@mail.ustc.edu.cn}}

\maketitle              

\input{sections/abstract_v1}

\section{Introduction}
\input{sections/introduction_v5}
\label{sec:intro}
\begin{figure}[ht]
   \begin{center}
   \includegraphics[width=\textwidth]{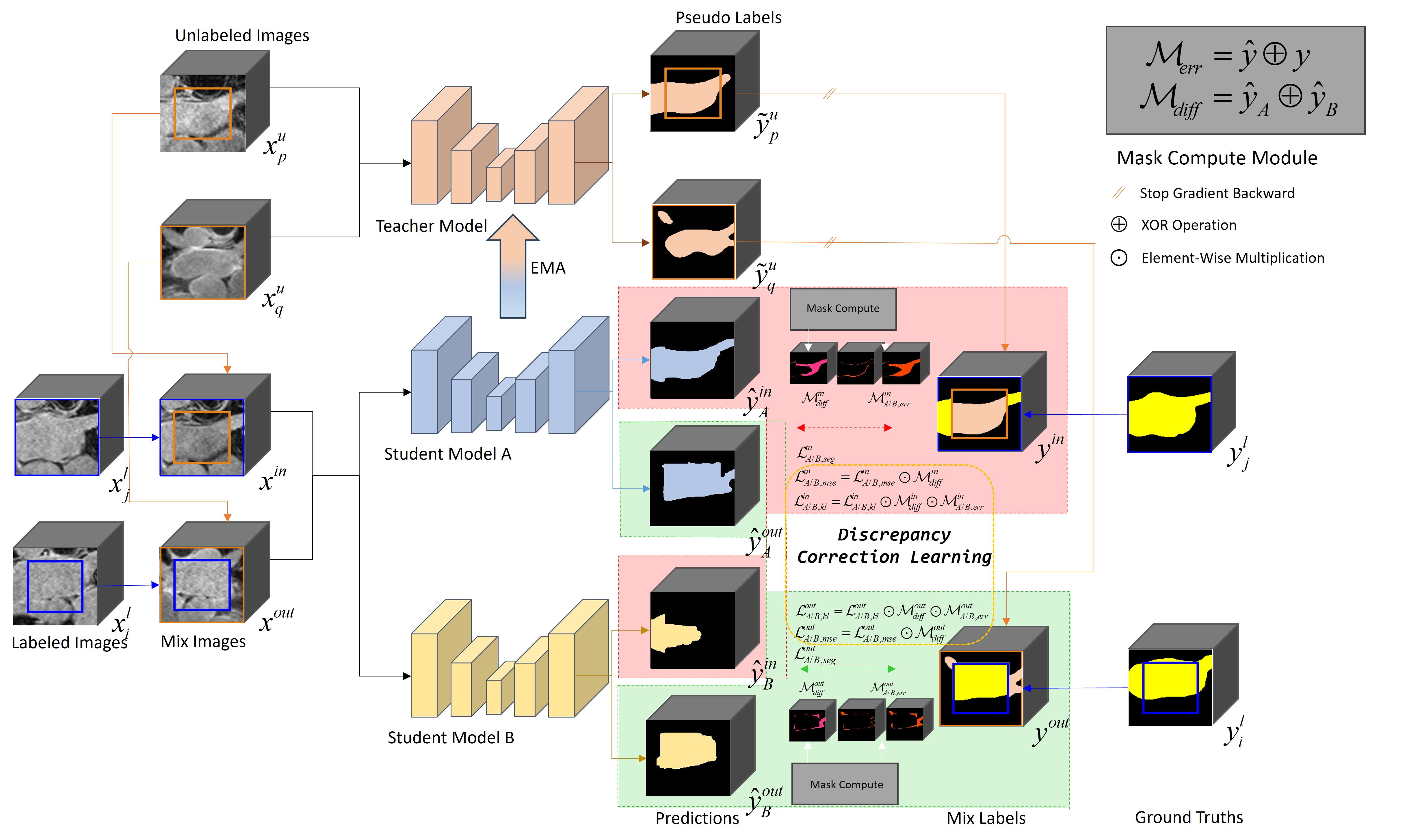}
   \caption{
    Overview of the proposed students discrepancy-informed correction learning (SDCL) framework for 
    semi-supervised medical image segmentation. 
    The SDCL framework adopts a BCP strategy, merging two labeled and two unlabeled images to create mix images. The teacher uses 
    unlabeled images to generate pseudo-labels. Ground truths and pseudo-labels are then mixed to 
    produce mix labels. Students A and B process the mixed images separately to calculate 
    segmentation loss. Finally, mask computation determines discrepancy and error masks, guiding the correction learning process.
   }
   \label{fig1}
   \end{center}
\end{figure}

\section{Methodology}
\input{sections/method_v1}
\label{sec:method}
\begin{table}[ht]
    \caption{Performance comparison on the Left Atrium and Pancreas datasets.}
    \begin{adjustbox}{max width=0.9\textwidth, center}
    \begin{tabular}{l|ccllll|ccllll}
        \hline
        Method    & \multicolumn{6}{l|}{Pancreas-CT}                                                                                                                                                & \multicolumn{6}{l}{Left  Atrium}                                                                                                                                               \\ \cline{2-13} 
                  & Lb  & \multicolumn{1}{c|}{Unlb} & Dice                  & Jac                   & 95HD               & ASD                & Lb & \multicolumn{1}{c|}{Unlb} & Dice                  & Jac                   & 95HD               & ASD                \\ \hline
        V-Net     & 12 & \multicolumn{1}{c|}{0}         & 70.59                           & 56.77                           & 14.19                          & 2.25                           & 8 & \multicolumn{1}{c|}{0}         & 79.87                           & 67.60                           & 26.65                          & 7.94                           \\
        ResV-Net  & 12 & \multicolumn{1}{c|}{0}         & 68.94                           & 54.50                           & 13.86                          & 3.36                           & 8 & \multicolumn{1}{c|}{0}         & 80.07                           & 69.29                           & 19.50                          & 6.02                           \\
        V-Net     & 62  & \multicolumn{1}{c|}{0}         & 82.60                           & 70.81                           & 5.61                           & 1.33                           & 80 & \multicolumn{1}{c|}{0}         & 91.47                           & 84.36                           & 5.48                           & 1.51                           \\
        ResV-Net  & 62  & \multicolumn{1}{c|}{0}         & 82.46                           & 70.50                           & 5.45                           & 1.44                           & 80 & \multicolumn{1}{c|}{0}         & 91.09                           & 83.90                           & 4.77                           & 1.75                           \\ \hline
        UA-MT~\cite{yu2019uncertainty}(MICCAI'19)    &          & \multicolumn{1}{c|}{}          & 77.26                           & 63.82                           & 11.90                          & 3.06                           &         & \multicolumn{1}{c|}{}          & 87.79                           & 78.39                           & 8.68                           & 2.12                           \\
        DTC~\cite{luo2021semi}(AAAI'21)       &          & \multicolumn{1}{c|}{}          & 78.27                           & 64.75                           & 8.36                           & 2.25                           &         & \multicolumn{1}{c|}{}          & 87.51                           & 78.17                           & 8.23                           & 2.36                           \\
        CoraNet~\cite{shi2021inconsistency}(TMI'21)   &          & \multicolumn{1}{c|}{}          & 79.67                           & 66.69                           & 7.59                           & 1.89                           &         & \multicolumn{1}{c|}{}          & -                               & -                               & -                              & -                              \\
        SS-Net~\cite{wu2022exploring}(MICCAI'22)    &  & \multicolumn{1}{c|}{}  & -                               & -                               & -                              & -                              &  & \multicolumn{1}{c|}{}  & 88.55                           & 79.62                           & 7.49                           & 1.90                           \\
        MC-Net+~\cite{wu2022mutual}(MIA'22)   &          & \multicolumn{1}{c|}{}          & 80.59                           & 68.08                           & 6.47                           & 1.74                           &         & \multicolumn{1}{c|}{}          & 88.96                           & 80.25                           & 7.93                           & 1.86                           \\
        CAML~\cite{gao2023correlation}(MICCAI'23)     &    12      & \multicolumn{1}{c|}{50}          & -                               & -                               & -                              & -                              &    8     & \multicolumn{1}{c|}{72}          & 89.62                           & 81.28                           & 8.76                           & 2.02                           \\
        DMD~\cite{xie2023deep}(MICCAI'23)      &          & \multicolumn{1}{c|}{}          & -                               & -                               & -                              & -                              &         & \multicolumn{1}{c|}{}          & 89.70                           & 81.42                           & 6.88                           & 1.78                           \\
        MCCauSSL~\cite{miao2023caussl}(ICCV'23)      &          & \multicolumn{1}{c|}{}          & 80.92                               & 68.26                               & 8.11                              &1.53                              &         & \multicolumn{1}{c|}{}          & -                           & -                           & -                           & -                           \\
        UPCoL~\cite{lu2023upcol}(MICCAI'23)      &          & \multicolumn{1}{c|}{}          & 81.78                               & 69.66                               & 3.78                              & 0.63                              &         & \multicolumn{1}{c|}{}          & -                           & -                           & -                           & -                           \\
        BCP~\cite{ref2}(CVPR'23)       &          & \multicolumn{1}{c|}{}          & 82.91                           & 70.97                           & 6.43                           & 2.25                           &         & \multicolumn{1}{c|}{}          & 89.62                           & 81.31                           & 6.81                           & 1.76                           \\
        SDCL(Ours) &          & \multicolumn{1}{c|}{}          & \textbf{85.04} & \textbf{74.22} & \textbf{5.22} & \textbf{1.48} &         & \multicolumn{1}{c|}{}          & \textbf{92.35} & \textbf{85.83} & \textbf{4.22} & \textbf{1.44} \\ \hline
        \end{tabular}
    \end{adjustbox}
    \label{tab:tab1}
\end{table}
\begin{table}[ht]
    \caption{Performance on ACDC dataset}
    \begin{adjustbox}{max width=0.9\textwidth, center}

    \begin{tabular}{l|ccllll}
        \hline
        Method                                            & \multicolumn{6}{l}{ACDC}                                                                                                                                             \\ \cline{2-7} 
                                                          & Lb & \multicolumn{1}{c|}{Unlb} & Dice                            & Jac                             & 95HD                           & ASD                            \\ \hline
        U-Net                                             & 7  & \multicolumn{1}{c|}{0}    & 79.41                           & 68.11                           & 9.35                           & 2.70                           \\
        ResU-Net                                          & 7  & \multicolumn{1}{c|}{0}    & 80.04                           & 68.73                           & 7.83                           & 1.94                           \\
        U-Net                                             & 70 & \multicolumn{1}{c|}{0}    & 91.65                           & 84.93                           & 1.89                           & 0.56                           \\
        ResU-Net                                          & 70 & \multicolumn{1}{c|}{0}    & 90.44                           & 82.95                           & 1.77                           & 0.47                           \\ \hline
        UA-MT~\cite{yu2019uncertainty}{(MICCAI'19)}   &    & \multicolumn{1}{c|}{}     & 81.65                           & 70.64                           & 6.88                           & 2.02                           \\
        SASSNet~\cite{li2020shape}{(MICCAI'20)} &    & \multicolumn{1}{c|}{}     & 84.50                           & 74.34                           & 5.42                           & 1.86                           \\
        DTC~\cite{luo2021semi}{(AAAI'21)}       &    & \multicolumn{1}{c|}{}     & 84.29                           & 73.92                           & 12.81                          & 4.01                           \\
        SS-Net~\cite{wu2022exploring}{(MICCAI'22)}  &    & \multicolumn{1}{c|}{}     & 86.78                           & 77.67                           & 6.07                           & 1.40                           \\
        MC-Net+~\cite{wu2022mutual}{(MIA'22)}    & 7  & \multicolumn{1}{c|}{63}   & 87.10                           & 78.06                           & 6.68                           & 2.00                           \\
        DC-Net~\cite{chen2023decoupled}{(MICCAI'23)}  &    & \multicolumn{1}{c|}{}     & 89.42                           & 81.37                           & 1.28                           & 0.38                           \\
        BCPCauSSL~\cite{miao2023caussl}{(ICCV'23)} &    & \multicolumn{1}{c|}{}     & 89.66                           & 81.79                           & 3.67                           & 0.93                           \\
        BCP~\cite{ref2}{(CVPR'23)}       &    & \multicolumn{1}{c|}{}     & 88.84                           & 80.62                           & 3.98                           & 1.17                           \\
        SDCL{(Ours)}          &    & \multicolumn{1}{c|}{}     & \textbf{90.92} & \textbf{83.83} & \textbf{1.29} & \textbf{0.34} \\ \hline
        \end{tabular}
    \end{adjustbox}
    \label{tab:tab2}
\end{table}
\section{Experiments}
\input{sections/experiment_v1}
\label{sec:experiment}

\section{Conclusion}

\input{sections/conclusion_v1}

\label{sec:conclusion}

\bibliographystyle{splncs04}
\bibliography{main.bib}

\end{document}

%% file: sections/abstract_v1.tex
\begin{abstract}

Semi-supervised medical image segmentation (SSMIS) has been demonstrated the potential to mitigate the issue of 
limited medical labeled data. However, confirmation and cognitive biases may affect the prevalent teacher-student based SSMIS methods due to erroneous pseudo-labels. 
To tackle this challenge, we improve the mean teacher approach and propose the Students Discrepancy-Informed Correction Learning 
(SDCL) framework that includes two students and one non-trainable teacher, which utilizes the segmentation difference between the two students to guide the self-correcting learning. 
The essence of SDCL is to identify the areas of 
segmentation discrepancy as the potential bias areas, and then encourage the model to review the correct cognition and rectify 
their own biases in these areas.  
To facilitate the bias correction learning with continuous review and rectification, two correction loss functions are employed to minimize the correct segmentation voxel distance and maximize the erroneous segmentation voxel entropy.  
We conducted experiments on three public medical image datasets: 
two 3D datasets (CT and MRI) and one 2D dataset (MRI). The results show that our SDCL surpasses 
the current State-of-the-Art (SOTA) methods by 2.57\%, 3.04\%, and 2.34\% in the Dice score on 
the Pancreas, LA, and ACDC datasets, respectively. 
In addition, the accuracy of our method is very close to the fully supervised method on the ACDC dataset, 
and even exceeds the fully supervised method on the Pancreas and LA dataset. 
(Code available at \url{https://github.com/pascalcpp/SDCL}).

\keywords{Semi-supervised learning \and Medical image segmentation \and Correction learning}
\end{abstract}

%% file: sections/introduction_v5.tex
Deep learning techniques have revolutionized medical image processing, mainly due to the superiority of neural network 
algorithms over many conventional image processing techniques. 
Relative to the general computer vision field, medical image segmentation 
encounters the dual challenges of the scarcity of annotation data and the 
greater complexity of datasets. These data constraints suppress the accuracy 
of medical image segmentation, and semi-supervised learning (SSL) methods are increasingly 
being recognized for their potential to tackle these challenges by leveraging both labeled and unlabeled data.

Current popular SSMIS methods focus on self-training, uncertainty estimation, consistency regularization, and distribution alignment. 
Self-training methods, such as self-training and co-training~\cite{zhou2019semi, bai2017semi}, use current high-confidence 
pseudo-labels and ground truths for iteratively training. Uncertainty estimation uses measures like information entropy to assess unlabeled data and guide pseudo-label filtering or weighting. Mean Teacher (MT)~\cite{ref1} 
framework is a prevalent approach that enforces consistency between student and teacher models. UA-MT~\cite{yu2019uncertainty} refines consistency learning of MT with uncertainty, 
while CoraNet~\cite{shi2021inconsistency} applies different weights to teacher-generated pseudo-labels based on uncertainty. 
BCP~\cite{ref2} aims to reduce the empirical distribution gap by learning common semantics from both labeled and unlabeled data in the MT framework.

Despite the ongoing progress in SSMIS, confirmation and cognitive biases remains a critical limitation, 
especially in the widely used semi-supervised learning based on teacher-student framework. 
Since the framework often adds input perturbations and applies consistency regularization between teacher and student, 
a single model structure inevitably produces noisy or erroneous pseudo-labels~\cite{ref2,ref4}, resulting in model confirmation and cognitive biases~\cite{arazo2020pseudo,ref4}. 
These biases limit the performance of the teacher-student framework and it is very difficult for the model to correct these biases on its own.

Recently, methods like multi-student~\cite{chen2021semi, luo20222semi} and multi-teacher~\cite{liu2022perturbed, na2024switching} have emerged to provide diverse pseudo-labels 
to mitigate confirmation and cognitive biases. Multi-student approaches involve cross-consistency learning, 
but they may encounter training instability without an Exponential Moving Average (EMA) teacher, 
and diversity and stability can be diﬀicult to balance. On the other hand, 
the multi-teacher methods employ a single student to update multiple teachers with various strategies 
to promote diverse learning. However, these methods are constrained by the single model structure, 
limiting their ability to adequately address biases. 
The researchers have also employed correction learning methods, such as~\cite{zhang2021self} incorporating dual-task network for bias correction and~\cite{wu2019mutual} utilizing complementary network for mapping to ground truth. MCF~\cite{ref4} proposes inter-subnet interaction as a means of bias corrections.

In this view, the teacher-student methods still lack a general approach for 
stablely rectify own biases using diverse information, and the incorporation of bias 
correction can help to improve the performance of SSMIS. Therefore, we propose 
students discrepancy-informed correction learning (SDCL) based on the Mean Teacher (MT) framework, 
featuring one self-ensembling teacher with two trainable students. We ensure stability with an 
EMA teacher and promote diversity by using students with different structures. SDCL 
considers the discrepancy areas between the segmentations of the two students as the 
potential bias areas, and then conducts correction learning in these areas. The contributions of this study include: 
(1) Different from the traditional teacher-student framework, we use two structurally different students and an EMA teacher to ensure the diversity and stability of the teacher-student framework.   
(2) We design a method to optimize bias correction learning that reviews correct cognition and rectifies error biases in the differences between the predictions of two students. 
(3) Our approach outperforms SOTA SSMIS methods on three datasets, and additionally, it performs comparably or surpasses the fully supervised method.

%% file: sections/method_v1.tex
\subsection{Problem Definition}
Given a medical image dataset $\mathcal{D}$, it contains $N$ labeled images $\mathcal{D}^l$ and $M$ 
unlabeled images $\mathcal{D}^u$ $(N \ll M)$, i.e. 
$\mathcal{D}=\left \{ \mathcal{D}^l,\mathcal{D}^u \right \}$, 
where 
$\mathcal{D}^l= \{ (x^l_i,y^l_i) \}_{i=1}^{N}$ and $\mathcal{D}^u=\left \{ x^u_i \right \}_{i=N}^{N+M}$. 
Each 3D volume medical image $x_i \in R^{W \times H \times D}$  
in $\mathcal{D}^l$ have label $y^l_i \in {\{0, 1, ..., K - 1\}}^{W \times H \times D}$. 
The output prediction of the model is 
$\hat y_i \in \{0, 1, ..., K - 1\}^{W \times H \times D}$.
SDCL includes two students and a self-ensembling teacher.

\subsection{SDCL Framework}
Our framework consists of two phases: initial pre-training with labeled data using Copy-Paste augmentation~\cite{ref2,ref3}, followed by a semi-supervised learning (SSL) phase incorporating both labeled and unlabeled data. SSL begins by initializing students and the teacher with the pre-trained model. 
The process of SSL includes three parts: 
i) obtaining the basic SSL segmentation losses based on the BCP strategy, 
ii) obtaining DiffMask 
$\mathcal{M}_{diff}$ from student segmentation discrepancies to guide the framework in reviewing correct cognition voxels, and iii) generating ErrMask 
$\mathcal{M}_{err}$ from the difference between student segmentations and mix labels, then creating DiffErrMask 
$\mathcal{M}_{differr}$ by multiplying $\mathcal{M}_{diff}$ and 
$\mathcal{M}_{err}$, which guides the repair of self-bias error voxels. 
To achieve diversity between the two students, 
for 3D tasks, we employ VNet as student A 
and ResVNet~\cite{ref4} as student B. Meanwhile, 
for 2D tasks, UNet is designated as student A, 
and ResUNet is assigned as student B. 
To accurately evaluate the impact of discrepancy correction learning, efforts are made to minimize interference from other factors.
We use VNet/UNet (student A) for Exponential 
Moving Average (EMA) updates to the teacher, 
aligning with other methodologies. 
It's important to emphasize that the 
performance gap between the two students is minimal. 
The framework is illustrated in Fig.~\ref{fig1}.

\subsection{Bidirectional Copy-Paste}
We combine our framework with the current SSMIS SOTA method BCP~\cite{ref2}. 
In the SSL phase, we need to generate a zero-centered 
mask $\mathcal{M} \in {(0, 1)}^{W \times H \times D}$,
where 0 represents foreground and 1 represents background, the size of 
0 region is $\beta W \times \beta H \times \beta D$, $\beta \in (0,1)$. Next, 
we use the mask to obtain mix images as the input of SSL as follows:
\begin{equation}
    x^{in} = x^l_j \odot \mathcal{M} + x^u_p \odot (1-\mathcal{M}),
    \qquad
    x^{out} = x^u_q \odot \mathcal{M} + x^l_i \odot (1-\mathcal{M}).
\end{equation}
To obtain the pseudo labels, 
$x^u_p$ and $x^u_q$ are forwarded to the teacher to obtain pseudo-labels $p^u_p$ and $p^u_q$. Because 
pseudo-labels contain a lot of noise~\cite{ref2,ref5}, which is very 
 harmful to model training, the optimized 
 pseudo labels $\tilde{y}^u_p$ and $\tilde{y}^u_q$ are obtained by selecting the 
 largest connected component of raw pseudo labels.
 Then the mix labels are defined as follows:
\begin{equation}
    y^{in} = y^l_j \odot \mathcal{M} + \tilde{y}^u_p \odot (1-\mathcal{M}),
    \qquad
    y^{out} = \tilde{y}^u_q \odot \mathcal{M} + y^l_i \odot (1-\mathcal{M}),
\end{equation}
where $i \neq j$ and $p \neq q$, and $ \odot $ denotes element-wise multiplication.
Next, fed $x^{in}$ and $x^{out}$ into two students to obtain $\hat y^{in}$ and $\hat y^{out}$ for each student. 
Finally, we get BCP losses computed respectively by Eq. (\ref{eq:seg1}) and Eq. (\ref{eq:seg2}):
\begin{equation}
    \mathcal{L}^{in}_{A/B, seg} = \mathcal{L}_{seg}(\hat y^{in}_{A/B}, y^{in})
    \odot \mathcal{M} + \alpha \mathcal{L}_{seg}(\hat y^{in}_{A/B}, y^{in})
     \odot (1 - \mathcal{M}),
     \label{eq:seg1}
\end{equation}
\begin{equation}
     \mathcal{L}^{out}_{A/B, seg} = \mathcal{L}_{seg}(\hat y^{out}_{A/B}, y^{out}) 
     \odot (1 - \mathcal{M}) + \alpha \mathcal{L}_{seg}(\hat y^{out}_{A/B}, y^{out})
      \odot \mathcal{M},
      \label{eq:seg2}
\end{equation}
$\mathcal{L}_{seg}$ consists of Dice and Cross-entropy loss in equal proportions. Since ground truths are generally more accurate than pseudo labels, 
$\alpha$ is used to control the weight between them.

\subsection{Discrepancy Correction Learning}
\subsubsection{Minimize Discrepancy Correct Distance.}
From an intuitively straightforward perspective, during training, 
we can increase the weighting of learning in the correct regions to 
review correct voxels and avoid model biases. Here, we minimize Mean Squared Error (MSE) to enhance the model's learning towards the correct voxels in discrepant regions. 
We first apply argmax to $\hat y^{in}$ and $\hat y^{out}$, resulting in $\tilde y^{in}$ and $\tilde y^{out}$. Then, $\mathcal{M}_{diff}$ is then derived from the following formula:
    $\mathcal{M}_{diff}^{in/out} = \tilde y^{in/out}_A \oplus \tilde y^{in/out}_B,$
where A and B respectively represent student A and B, $\oplus$ denotes XOR operation. 
Intuitively, we can obtain losses using a method similar to BCP
, and the MSE losses are computed through Eq. (\ref{eq:mse1}), Eq. (\ref{eq:mse2}), and Eq. (\ref{eq:mse3}):
\begin{equation}
    \mathcal{L}^{in}_{A/B, mse} = \mathcal{L}_{mse}(\hat y^{in}_{A/B}, y^{in})
    \odot \mathcal{M} + \alpha \mathcal{L}_{mse}(\hat y^{in}_{A/B}, y^{in})
     \odot (1 - \mathcal{M}),
     \label{eq:mse1}
\end{equation}
\begin{equation}
     \mathcal{L}^{out}_{A/B, mse} = \mathcal{L}_{mse}(\hat y^{out}_{A/B}, y^{out}) 
     \odot (1 - \mathcal{M}) + \alpha \mathcal{L}_{mse}(\hat y^{out}_{A/B}, y^{out})
      \odot \mathcal{M},
      \label{eq:mse2}
\end{equation}
\begin{equation}
    \mathcal{L}^{in/out}_{A/B, mse} = \mathcal{L}^{in/out}_{A/B, mse} \odot \mathcal{M}_{diff}^{in/out}.
    \label{eq:mse3}
\end{equation}
\subsubsection{Maximize Discrepancy Erroneous Entropy.}
In the teacher-student framework, inherent confirmation and cognitive biases emerge. We incorporate a loss function for penalization to enable the model to self-correct these biases. Maximizing the entropy of erroneous voxels may be an effective strategy, pulling misclassified voxels in uncertain regions back to an initial state and redirecting them toward the correct direction.
The entropy of each voxel is defined as $\mathcal{H}(\hat y^{(x,y,z)}) = -\sum_{c=0}^{K-1} \hat y^{(x,y,z)}(c) \log{\hat y^{(x,y,z)}(c)}$, where $(x,y,z)$ denotes the voxel's position, and $\hat y^{(x,y,z)} \in \mathcal{R}^K$. The objective is to maximize $\mathcal{H}(\hat y^{(x,y,z)})$ for each erroneously classified voxel. We employ Kullback-Leibler (KL) divergence~\cite{belharbi2021deep}, a simple equivalent variant, to guide misclassified voxels in shifting their output distribution towards a uniform distribution. This minimizes the loss equation $\mathcal{L}_{kl}(\hat y, u) = \mathcal{D}_{KL}(u||\hat y)$, where $u$ represents the uniform distribution that all components are equal to $\frac{1}{K}$.
Before deriving losses, we obtain $\mathcal{M}_{err}$ using the following expression:
$\mathcal{M}_{A/B, err}^{in/out} = \tilde y^{in/out}_{A/B} \oplus y^{in/out}.$
Afterward, $\mathcal{M}^{in/out}_{A/B, differr} = \mathcal{M}^{in/out}_{diff} \odot \mathcal{M}^{in/out}_{A/B, err}$. 
The KL losses we obtain are as shown in the following equations Eq. (\ref{eq:kl1}), Eq. (\ref{eq:kl2}), and Eq. (\ref{eq:kl3}): 
\begin{equation}
    \mathcal{L}^{in}_{A/B, kl} = \mathcal{L}_{kl}(\hat y^{in}_{A/B}, u)
    \odot \mathcal{M} + \alpha \mathcal{L}_{kl}(\hat y^{in}_{A/B}, u)
     \odot (1 - \mathcal{M}),
     \label{eq:kl1}
\end{equation}
\begin{equation}
     \mathcal{L}^{out}_{A/B, kl} = \mathcal{L}_{kl}(\hat y^{out}_{A/B}, u) 
     \odot (1 - \mathcal{M}) + \alpha \mathcal{L}_{kl}(\hat y^{out}_{A/B}, u)
      \odot \mathcal{M},
      \label{eq:kl2}
\end{equation}
\begin{equation}
    \mathcal{L}^{in/out}_{A/B, kl} = \mathcal{L}^{in/out}_{A/B, kl} \odot \mathcal{M}_{A/B, differr}^{in/out}.
    \label{eq:kl3}
\end{equation}
In the final step, we form the total loss 
by linearly combining $\mathcal{L}_{seg}$, 
$\mathcal{L}_{mse}$
and $\mathcal{L}_{kl}$ with specific weights, 
as shown in Eq. (\ref{eq:total}),
\begin{equation}
    \mathcal{L}_{A/B} = \mathcal{L}^{in}_{A/B, seg} + \mathcal{L}^{out}_{A/B, seg}
    + \newline
    \gamma (\mathcal{L}^{in}_{A/B, mse} + \mathcal{L}^{out}_{A/B, mse}) + 
    \mu (\mathcal{L}^{in}_{A/B, kl} + \mathcal{L}^{out}_{A/B, kl}).\label{eq:total}
\end{equation}

%% file: sections/experiment_v1.tex
\noindent \textbf{Datasets.} 
Our method is evaluated on three datasets: the Pancreas-NIH~\cite{roth2015deeporgan} with 82 CT volumes (12 labeled (20\%), 50 unlabeled), the LA (Left Atrium)~\cite{xiong2021global} dataset with 100 
3D GE-MRIs (8 labeled (10\%), 72 unlabeled for training, 20 for testing), and the ACDC~\cite{bernard2018deep} 
dataset with 100 cardiac MRI scans (70 for training, 10 for validation, 20 for testing, with 7 labeled(10\%) 
and 63 unlabeled). For fair comparison, we use the same experimental setup as prior works like CoraNet 
and SS-Net~\cite{shi2021inconsistency,wu2022exploring,luo2022semi}, normalizing images 
and applying standard data augmentation. We follow the dataset 
splits used in previous studies. The ACDC result represents the average performance of four-class segmentation on the test set.
\begin{figure}[ht]
    \begin{center}
    \includegraphics[width=\textwidth]{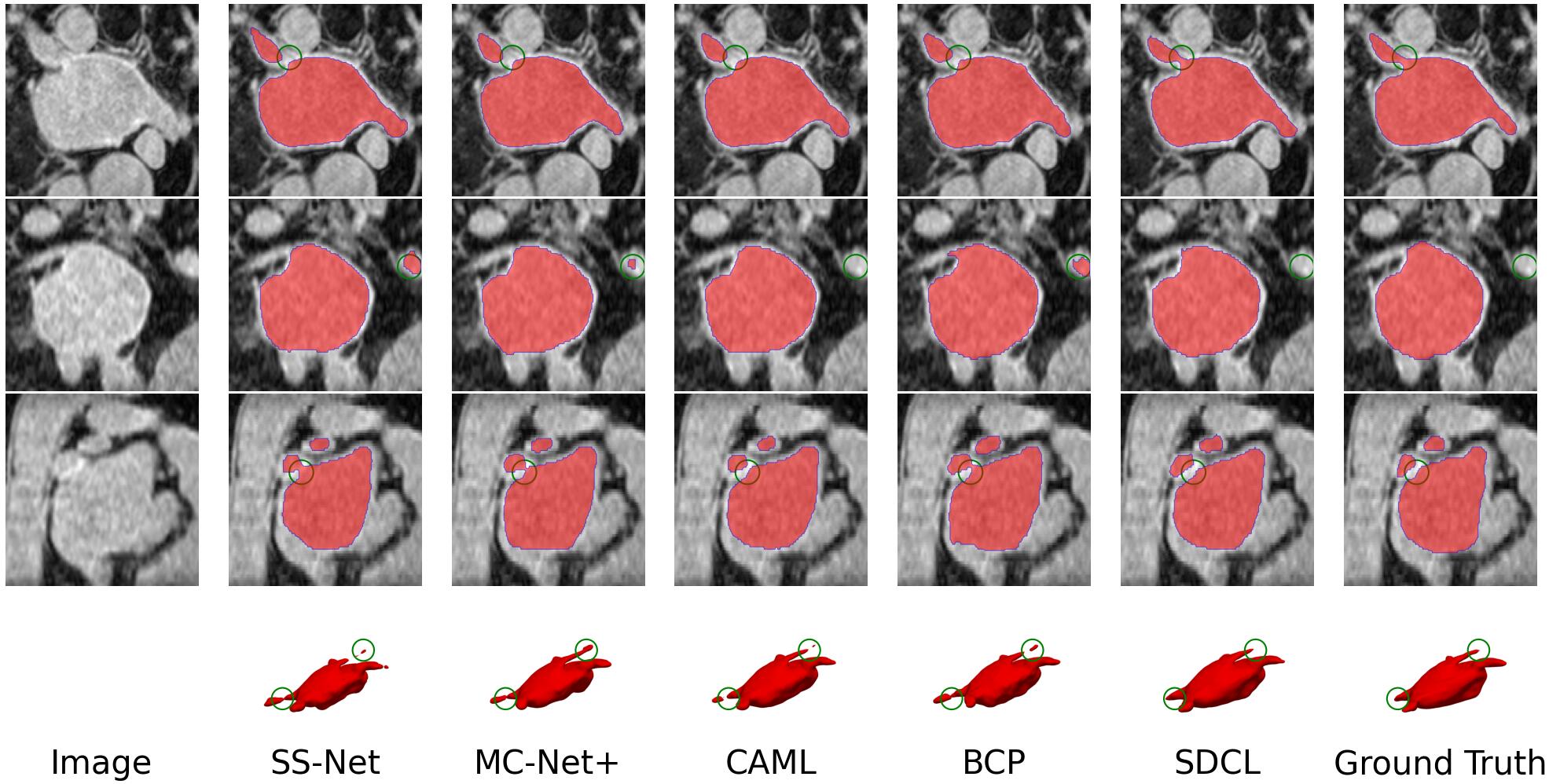}
    \caption{2D and 3D Visualization of segmentation results on Left Atrium dataset.}
    \label{fig2}
    \end{center}
\end{figure}

\noindent \textbf{Implementation Details.}
We used PyTorch 2.1.0 and an NVIDIA RTX 4090 GPU to experiment, averaging segmentation results from students A and B for evaluation. Pre-training had iterations of 3k (Pancreas), 3k (LA), and 11k (ACDC), while SSL had iterations of 8k (Pancreas), 10k (LA), and 45k (ACDC). We employed the Adam optimizer with a learning rate of 0.001, a 
batch size of 8, 8, 48 (half for labeled and half for unlabeled data), and set hyperparameters $\alpha=0.5$, 
$\beta=2/3$. Specific values for $\gamma$ and $\mu$ were 0.3 and 0.1 (Pancreas), 0.5 and 0.05 (LA and ACDC). 
Input patch sizes were $96 \times 96 \times 96$ (Pancreas), $112 \times 112 \times 80$ (LA), and $256 \times 256$ 
(ACDC). Testing used strides of $16 \times 16 \times 4$ (Pancreas) and $18 \times 18 \times 4$ (LA)~\cite{shi2021inconsistency,wu2022exploring,luo2022semi}, with 
no post-processing during evaluation. Performance metrics included Dice Score (Dice), Jaccard Score (Jac), 
95\% Hausdorff Distance (95HD), and Average Surface Distance (ASD).

\noindent \textbf{Results on the LA and Pancreas Datasets.}
Table~\ref{tab:tab1} provides a detailed comparison of our 
approach with the current SOTA SSMIS methods and presents the performance upper and lower bounds of our two backbones using only labeled data in a fully supervised manner. The table demonstrates significant improvements in our method across four metrics 
compared to the baseline (BCP) and other SOTA methods. As depicted in Fig.~\ref{fig2}, our approach closely aligns with the Ground Truth, especially in regions prone to errors at boundaries and connections. This underscores the efficacy of our discrepancy correction learning, significantly contributing to enhancing the model's edge and shape segmentation ability.

\noindent \textbf{Results on the ACDC Dataset.}
Table.~\ref{tab:tab2} presents a comparative analysis of our results against current methods and the upper and lower bounds of fully supervised methods. Similarly, our method exhibits a substantial improvement over the baseline, particularly evident in a significant enhancement of the ASD metric. Our approach demonstrated a 39\% reduction on the metric of ASD relative to the fully supervised upper bound of U-Net. The presence of more 2D slices in the dataset likely contributes to the enhancement of geometric segmentation integrity through discrepancy correction learning.

\noindent \textbf{Ablation Study.}
Table.~\ref{tab:tab3}, we conducted an ablation study on various components of our framework, comparing their impact on Dice scores against the baseline. 
Omitting $\mathcal{M}_{diff}$, $\mathcal{L}_{mse}$ 
and $\mathcal{L}_{kl}$ individually 
resulted in improvements of 0.43\% and 0.38\%, 
while the simultaneous use of both losses showed 
an enhancement of over 0.53\%. Introducing $\mathcal{M}_{diff}$ increased the improvements to 1.17\% and 1.26\%, and utilizing all components achieved a significant improvement of 2.16\%. These findings highlight the positive effect of incorporating correct cognition review and self-bias error correction in SDCL on model performance. 
Notably, we observed the most favorable outcomes when simultaneously employing both losses and $\mathcal{M}_{diff}$, affirming the efficacy of our proposed correction learning based on students' discrepancy.
For further insights, the Supplementary Materials include the variation of biased error voxels during the training process, detailed experiment results, and hyperparameter ablation studies in this paper.
\begin{table}[ht]
    \caption{Ablation study results on the pancreas dataset.}
    \begin{adjustbox}{max width=1\textwidth, center}
    \begin{tabular}{cc|llll|llll}
        \hline
        \multicolumn{2}{l|}{Scans used} & \multicolumn{4}{l|}{Components}                                              & \multicolumn{4}{l}{Metrics}                   \\ \hline
        Lb        & Unlb      & \multicolumn{1}{l|}{$\mathcal{L}_{seg}$} & \multicolumn{1}{l|}{$\mathcal{L}_{mse}$} & \multicolumn{1}{l|}{$\mathcal{L}_{kl}$} & $\mathcal{M}_{diff}$ & Dice & Jac & 95HD & ASD \\ \hline
                        &                & \multicolumn{1}{l|}{$\sqrt{}$} & \multicolumn{1}{l|}{}  & \multicolumn{1}{l|}{}  &   & 83.23        & 71.57      & 8.53           & 2.49          \\ \cline{3-10} 
                        &                & \multicolumn{1}{l|}{$\sqrt{}$} & \multicolumn{1}{l|}{$\sqrt{}$} & \multicolumn{1}{l|}{}  &   & 83.59        & 72.18        & 7.20           & 2.30          \\
                        &                & \multicolumn{1}{l|}{$\sqrt{}$} & \multicolumn{1}{l|}{$\sqrt{}$} & \multicolumn{1}{l|}{}  & $\sqrt{}$ & 84.20        & 73.01       & 6.25           & 2.03          \\ \cline{3-10} 
        12       & 50       & \multicolumn{1}{l|}{$\sqrt{}$} & \multicolumn{1}{l|}{}  & \multicolumn{1}{l|}{$\sqrt{}$} &   & 83.55        & 72.05       & 7.36           & 2.09          \\
                        &                & \multicolumn{1}{l|}{$\sqrt{}$} & \multicolumn{1}{l|}{}  & \multicolumn{1}{l|}{$\sqrt{}$} & $\sqrt{}$ & 84.28        & 73.12       & 6.31           & 1.97          \\ \cline{3-10} 
                        &                & \multicolumn{1}{l|}{$\sqrt{}$} & \multicolumn{1}{l|}{$\sqrt{}$} & \multicolumn{1}{l|}{$\sqrt{}$} &   & 83.67        & 72.20       & 9.12           & 2.80          \\
                        &                & \multicolumn{1}{l|}{$\sqrt{}$} & \multicolumn{1}{l|}{$\sqrt{}$} & \multicolumn{1}{l|}{$\sqrt{}$} & $\sqrt{}$ & 85.04        & 74.23       & 5.22           & 1.48         \\ \hline
    \end{tabular}
    \end{adjustbox}
    \label{tab:tab3}
\end{table}

%% file: sections/conclusion_v1.tex
We propose a novel SSMIS framework, extending the Mean-Teacher with an additional student for correction learning based on student discrepancies. The method aims to review correct voxels and repair self-bias error voxels in discrepancies and is compatible with other teacher-student models. It outperforms existing methods in 2D and 3D tasks. 
Future research will consider using students' information to refine teacher and further improve the performance of semi-supervised learning in medical image segmentation.
\begin{credits}
\subsubsection{\ackname}
This work was supported by the Natural Science Foundation of SiChuan, China (No.2022NSFSC0940).
\end{credits}